\documentclass[prd,letterpaper,tightenlines,nofootinbib,twocolumn,floatfix,showpacs]{revtex4}
\usepackage{amsmath,amssymb,graphicx,color}

\newcommand{\be}{\begin{equation}}
\newcommand{\bl}[1]{\begin{equation}\label{#1}}
\newcommand{\ee}{\end{equation}}
\newcommand{\pd}[2]{\frac{\partial{#1}}{\partial{#2}}}
\newcommand{\z}[1]{\left({#1}\right)}
\newcommand{\sz}[1]{\left[{#1}\right]}
\newcommand{\kz}[1]{\left\{{#1}\right\}}
\newcommand{\rec}[1]{\frac{1}{#1}}
\newcommand{\m}[1]{\mathrm{#1}}
\renewcommand{\c}[1]{\mathcal{#1}}
\renewcommand{\v}[1]{\mathbf{#1}}
\renewcommand{\r}[1]{(\ref{#1})}
\newcommand{\Eq}[1]{Eq.\@ (\ref{#1})}
\newcommand{\Eqs}[2]{Eqs.\@ (\ref{#1}) and (\ref{#2})}
\newcommand{\kfki}{MTA KFKI RMKI, H-1525 Budapest 114, POBox 49, Hungary}

\begin{document}

\title{New simple explicit solutions of perfect fluid hydrodynamics and phase-space evolution}
\author{M.~I.~Nagy}
\affiliation{\kfki}

\begin{abstract}
New exact solutions of relativistic perfect fluid hydrodynamics are described, including the first family of
exact rotating solutions. The method used to search for them is an investigation of the relativistic hydrodynamical
equations and the collisionless Boltzmann equation. Possible connections to the evolution of hot and dense partonic
matter in heavy-ion collisions is discussed.
\end{abstract}

\pacs{24.10.Nz,47.15.Hg}
\maketitle

\section{Introduction}

Hydrodynamical models are widely used in the description of collective properties in high-energy collisions:
in a thermal picture, the observed hadronic final state is treated as a result of particle production from a
thermal ensemble, and hydrodynamics is a powerful tool to dynamically connect this final state to the initial
state of a high energy collision. The solutions of the hydrodynamical equations are thus important because they
can shed light on the dynamics of the strongly interacting matter. Exact solutions\footnote{
We make the distinction between \emph{numerical} solutions (whose advantage is generality and applicability
to many different initial conditions) and \emph{exact} ones, formulated in terms of explicit formulas, the
advantage of them being the parametrization of a set of initial conditions and simplicity.
}
are important also theoretically: they are solutions of a highly coupled set of nonlinear differential equations.
They can be used to test numerical codes, and in some cases, they by themselves provide insight into the
dynamics of the hot and dense matter.

The application of hydrodynamics in the field of high-energy physics has a long history. In the '50s, Landau
and his collaborators formulated relativistic hydrodynamics, and gave a paradigmatic example of exact solutions,
nowadays known as the Landau-Khalatnikov solution~\cite{Landau:1953gs,Khalatnikov,Belenkij:1956cd}. Another
historic exact solution is the one-dimensional, boost-invariant accelerationless one, known as the Hwa-Bjorken
solution~\cite{Hwa:1974gn,Bjorken:1982qr}. In the past years, the heavy-ion experiments at the RHIC particle
accelerator yielded many surprising results on the the soft kinematic domain of particle production in heavy
ion collisions. The scaling behavior in the single-particle spectra, the two-particle correlations and anisotropy
(flow) properties could be explained under the assumption that the created matter is a strongly interacting
liquid. Because of this, the interest in relativistic hydrodynamics flourished again in the past years.

Because of their analytic simplicity, exact solutions of relativistic hydrodynamics are very useful in the
study of the space-time picture of high-energy reactions. The Landau-Khalatnikov solution was used for decades
in the description of cosmic ray elementary particle collisions. The Hwa-Bjorken solution yields an estimate
of the initial energy density of high-energy collisions~\cite{Bjorken:1982qr}. (Ref.~\cite{Csorgo:2006ax}
presents an improvement of this estimation.) Recently, an interpolating solution between the Landau-Khalatnikov
solution and the Hwa-Bjorken solution was discovered~\cite{Bialas:2007iu}.) The Buda-Lund model
\cite{Csanad:2004mm,Csanad:2005qr}, which gives a good description of hadronic observables, e.g. universal
scaling of elliptic flow, relies on a class of exact solutions~\cite{Csorgo:2001ru}. There are many other
exact solutions with important applications, see e.g. Refs.~\cite{Bondorf:1978kz,Csizmadia:1998ef,Csorgo:2001ru}
for some non-relativistic, Refs. \cite{Biro:2000nj,Csorgo:2003ry,Sinyukov:2004am,Bialas:2007iu,Borshch:2007uf,
Liao:2009zg,Nagy:2007xn,Csorgo:2006ax} for some relativistic examples.

In this paper the equations of relativistic hydrodynamics are approached from an unusual direction. There is a
non-relativistic exact solution~\cite{Csizmadia:1998ef}, described in detail in Section~\ref{s:notation}, that
has a remarkable property: the corresponding phase-space distribution of the flowing particles satisfies the
collisionless Boltzmann equation~\cite{Csizmadia:1998ef}. This means that one can prepare a microscopic initial
condition of the flowing particles, where the collisionless free motion of them maintains local thermalization,
and thus the macroscopic quantities (temperature, pressure, velocity) solve the hydrodynamical equations (since
they follow from kinetic theory). I will refer to this peculiarity as collisionless flow or \emph{Knudsen flow},
in analogy to the case of Knudsen gases, where the collisions of particles play no role in the thermalization.
(A hydrodynamical solution whose phase-space distribution satisfies the collisionless Boltzmann equation is
\emph{not} necessarily free of collisions, but it can be, for appropriate microscopic initial conditions.) 
The existence of such solutions may be surprising, because hydrodynamics relies on local thermal equilibrium, which
is usually maintained by collisions, so it is interesting that in some cases the particles behave thermodynamically
equilibrated even without collisions\footnote{The mentioned non-relativistic solution~\cite{Csizmadia:1998ef}
was discovered as a phase-space distribution whose general form is pertained with collisionless evolution, then it
was realized that this is a solution to the hydrodynamical equations~\cite{TCSpriv}.}.

In the following I investigate this idea and generalize it to relativistic hydrodynamics. The result is a new family
of exact solutions, even ones with nonzero rotation. (As far as I know, relativistic hydrodynamical solutions with nonzero
curl were not known before.) These and other space-time characteristics of the new solutions make them a good candidate for
using them in the description of high-energy reactions. 

\section{Notation and basic equations}\label{s:notation}

Let's denote the space-time coordinate by $x^\mu\equiv\z{t,\v{r}}$, and the metric tensor by
$g^{\mu\nu}\equiv diag\z{1,-1,\dots}$. We investigate $1+1$ and $1+3$ dimensional flows as well, the notation
$D$ will stand for the dimensionality of the space: $g^\mu_\mu=D+1$, $\delta_{kk}=D$. (Greek letters denote
Lorentz indices, Latin letters denote three-vector indices.) The four-velocity of the fluid is $u^\mu\equiv\gamma\z{1,\v{v}}$,
where $\gamma=\sqrt{1-v^2}$, and the thermodynamical quantities are: $T$ the temperature, $\varepsilon$ the energy density,
$p$ the pressure. (Their dependence on $x^\mu$ is usually suppressed in the notation.) The Equation of State (EoS) of the
matter connects $\varepsilon$ with $p$ and $T$. If we consider a fluid consisting of individual particles, then their
(conserved) number density is denoted by $n$, and the corresponding chemical potential by $\mu$. The fundamental
thermodynamical relations are
\bl{e:thermo}
\varepsilon+p=Ts+\mu n,\quad \m{d}p=s\m{d}T+n\m{d}\mu .
\ee
The hydrodynamical equations are well known, I recite them in the form used here. In the non-relativistic case, one
has to introduce the particle mass $m_0$. The Euler and the energy conservation equations are then
\bl{e:eul_nonrel}
nm_0\z{\pd{\v{v}}{t}+\z{\v{v}\nabla}\v{v}}=-\nabla p ,
\ee
\bl{e:energy_nonrel}
\pd{\varepsilon}{t}+\nabla\z{\varepsilon\v{v}}=-p\z{\nabla\v{v}} ,
\ee
and if a (conserved) particle number density is also considered, we have
\bl{e:cont_nonrel}
\pd{n}{t}+\nabla\z{n\v{v}}=0 .
\ee
The hydrodynamical equations need to be supplemented by an appropriate EoS to close the set of equations. Throughout
this paper, the following equation of state is used:
\bl{e:EoS}
\varepsilon=\kappa p ,\quad p=nT .
\ee
This EoS is used in most cases of exact solutions, eg. the Hwa-Bjorken solution, the Landau-Khalatnikov solution,
and many others. 
The constant $\kappa$ is usually left for free choice, but in this paper, all the solutions investigated are valid
only if we constrain the value of $\kappa$: we must have $\kappa=D/2$ in the nonrelativistic case, and $\kappa=D$
in the relativistic case\footnote{It must be noted that any hydrodynamical solution with $\varepsilon=\kappa p$
EoS has another important property: one can introduce the \emph{bag constant} $B$, and set $p\to p-B$,
$\varepsilon\to\varepsilon+B$ and the solutions still remain valid with the ,,bag'' equation of state, which is
$\varepsilon-B=\kappa\z{p+B}$.}. However, for clarity, $\kappa$ will be sometimes retained in the notation.

In the more interesting relativistic case, the Euler equation, and the energy and particle number conservation
equations are
\bl{e:eul_rel}
\z{\varepsilon+p}u^\nu\partial_\nu u^\mu=\z{g^{\mu\rho}-u^\mu u^\rho}\partial_\rho p , 
\ee
\bl{e:energy_rel}
\z{\varepsilon+p}\partial_\rho u^\rho+u^\rho\partial_\rho\varepsilon = 0 ,
\ee
\bl{e:n_rel}
\partial_\mu\z{nu^\mu}=0 .
\ee
If we do not investigate the $n$ density (and its continuity equation \r{e:n_rel}), then the equations contain
only $u^\mu$, $p$ and $\varepsilon$, so the $\varepsilon=\kappa p$ EoS closes the set of the equations, without
any reference to $T$ or $n$.

Nevertheless, a central topic of this paper, the relation between the collisionless phase-space evolution of the
matter and the hydrodynamical flow, is meaningful only if we include a non-vanishing, conserved $n$, and thus
$\mu$, and in the spirit of \Eq{e:EoS}, also $T$. The interesting quantity in kinetic theory, the phase-space
distribution $f$ is a function of the momentum $p^\mu=\z{E,\v{p}}$ and $x^\mu$. In the non-relativistic case, it
has the Maxwell-Boltzmann form as
\bl{e:f_nonrel}
f\z{t,\v{r},\v{p}}=\exp\z{\frac{\mu}{T}-\frac{\z{\v{p}-m_0\v{v}}^2}{2m_0T}} ,
\ee
while the relativistic generalization is
\bl{e:f_rel}
f\z{x,p} = \exp\z{\frac{\mu}{T}-\frac{p_\mu u^\mu}{T}} .
\ee
For non-vanishing $n$, $f$ is required to be normalized as 
\bl{e:norm}
\int\m{d}^Dp\,f\z{x,p}=n\z{x} .
\ee
(For example, in the non-relativistic case this is fulfilled if $\exp\z{\mu/T}=n\z{2\pi m_0 T}^{-D/2}$, this is one
of the reasons for choosing \Eq{e:EoS} to be the EoS, with the restrictions on $\kappa$ explained there.)

There is a known class of ellipsoidally symmetric non-relativistic exact solutions~\cite{Csorgo:2001ru}. A
spherically symmetric special case (mentioned in the introduction), with homogeneous temperature and Gaussian
density profile was discovered earlier~\cite{Csizmadia:1998ef}:
\[
\v{v}=\frac{\dot{a}\z{t}}{a\z{t}}\v{r} ,\quad n=n_0\z{\frac{a_0}{a\z{t}}}^3\exp\z{-\frac{r^2}{2a^2\z{t}}} ,
\]
\bl{e:sol_nonrel}
T=T_0\z{\frac{a_0}{a\z{t}}}^2 ,\quad a\z{t}=A\z{t-t_0}^2+B ,
\ee
with $a_0$, $t_0$, $A$, and $B$ constants. This solution has the surprising property explained in the
introduction: the collisionless phase-space evolution~\cite{Csizmadia:1998ef}. (With other words, Ref.~\cite{Csizmadia:1998ef} describes a Knudsen 
flow.) The Boltzmann equation without collisions takes a particularly simple form:
\bl{e:MB_nonrel}
\pd{}{t}f\z{\v{r},\v{p}}+\frac{\v{p}}{m_0}\pd{}{\v{r}}f\z{\v{r},\v{p}} = 0 .
\ee
Indeed, this has the general solution $f\z{t,\v{r},\v{p}}=f\z{t_0,\v{r}-\v{p}t/m_0,\v{p}}$, which is a free streaming.
The solution given by \Eq{e:sol_nonrel} satisfies this, because Gaussian distributions are stable, and the velocity
field is self-similar. Appendix \ref{s:app_nonrel} proves that there are no essentially different non-relativistic
collisionless solutions.

In the relativistic case, the collisionless Boltzmann-equation is 
\bl{e:MB_rel}
p^\mu\partial_\mu f=0 \quad\z{\forall p} ,
\ee
This is the relativistic analogue of \Eq{e:MB_nonrel}. In the next section the solutions of the coupled
(relativistic) hydrodynamical and collisionless Boltzmann equations are presented. The derivations are
rather technical, they are left for Appendix \ref{s:app_colless} and \ref{s:app_hydro}.

\section{New exact relativistic solutions and Knudsen flows}\label{s:hydro}

The equations \r{e:Amu} and \r{e:1Dsol} in Appendix \ref{s:app_colless} allow us to derive the collisionless
hydrodynamical solutions (the Knudsen flows). Some of them are special cases of known solutions, (here the new result is the
collisionlessness), some of them are new results in themselves. Some of these new solutions are special
cases of new more general solutions (which are not Knudsen flows). It must be noted that the new solutions are in a
sense independent of the investigation presented in the previous section: they are valid, exact,
explicit hydrodynamical solutions ($p$, $\varepsilon$ and $u^\mu$ fields) for the $\varepsilon=\kappa p$ ($\kappa=D$)
equation of state (as in \Eq{e:EoS}).

The 1+1 dimensional solution given by \Eq{e:1Dsol} (with $x\equiv x^1$ being the only spatial coordinate) can be
expressed with two arbitrary functions, $\chi\z{x^+}$ and $\xi\z{x^-}$ (with the notation $x^\pm\equiv t\pm x$) as
\bl{e:1DvT}
v=\frac{\chi\z{x^+}-\xi\z{x^-}}{\chi\z{x^+}+\xi\z{x^-}} ,\quad T=\frac{T_0/2}{\sqrt{\chi\z{x^+}\xi\z{x^-}}} .
\ee
Recall that we must have $\kappa=D$ in the EoS, that is, here $\kappa=D=1$. In Appendix D of Ref.~\cite{Nagy:2007xn},
a general solution of the hydrodynamical problem for $\kappa=D=1$ was presented, with similar characteristics: the
velocity field $v$ and the pressure $p$ is expressed as the combinations of two arbitrary wave-shapes propagating
in the opposite direction. (The notation used there is different, the $F$ and $G$ functions introduced there are
$F=\ln\chi$ and $G=-\ln\xi$, respectively). \Eq{e:1DvT} is a special case of that solution, when $p=nT$ and
$n=const\cdot T$, that is, $T=const\cdot\sqrt{p}$, which is a minor restriction on the general solution. We
see now that this restriction is enough to have a solution with collisionless phase-space evolution. 

We now turn to the three-dimensional case. In the following we denote $\left|\v{r}\right|$ by $r$, and make use
of the Rindler-like coordinates $\tau$ and $\eta$ (,,proper-time'' and ,,space-time pseudorapidity''), defined here as
\bl{e:Rindler}
\tau=\sqrt{t^2-r^2} ,\quad \eta=\rec{2}\ln\frac{t+r}{t-r} .
\ee
(Note that this notational convention favors spherical symmetry.) \Eq{e:Amu} in Appendix \ref{s:app_colless} offers
several possibilities for relativistic hydrodynamical solutions. The investigation of \Eq{e:Amu} is left to Appendix
\ref{s:app_hydro}, here the results are collected. The notations $T_0$, $p_0$ and $\tau_0$ always stand for
,,initial'' values (to set the scale of the temperature and the pressure).

The first case is the well-known Hwa-Bjorken solution in $1+1$ dimensions, and the Hubble-like (or Buda-Lund type)
solutions~\cite{Csorgo:2003ry} in $D\neq 1$:
\bl{e:BL}
\v{v}=\frac{\v{r}}{t},\quad T=T_0\frac{\tau_0}{\tau},\quad p=p_0\z{\frac{T}{T_0}}^{\kappa+1} ,
\ee
The solution itself is a known result, the new result is that it corresponds to a Knudsen flow. 

The second case of solutions is an accelerating solution, which is a generalization of the previously known solution in
Refs.~\cite{Nagy:2007xn,Csorgo:2006ax}. Its form is
\[
\v{v}=\frac{2t\v{r}+\v{a}}{t^2+r^2+\rho},\quad p=p_0\z{\frac{T}{T_0}}^{\kappa+1} ,
\]
\bl{e:NCC_mu}
T=\frac{T_0\tau_0^2}{\sqrt{\z{\tau^2+\rho}^2+4\rho\tau^2\sinh^2\eta-4\z{\v{a}\v{r}}t-a^2}} .
\ee
with arbitrary $\v{a}$ constant three-vector and $\rho\ge 0$ constant (the $\rho=0$, $\v{a}=0$ case was presented
in~\cite{Nagy:2007xn,Csorgo:2006ax}). For $\v{a}\neq 0$ this new solution prefers a direction in space. The
$\v{a}=0$, $\rho>0$ solution is sperically symmetric, and it is \emph{finite} in
$\eta$, what is an improvement on Refs.~\cite{Nagy:2007xn,Csorgo:2006ax}. (The term ,,finite'' means here that
$p$ drops to zero at any given $\tau=\tau_1$ hypersurface as $\eta$ goes to $\pm\infty$. Roughly speaking, the
domain in $\eta$ where the pressure and matter density is different from $0$ is finite.) 

\begin{figure}
\includegraphics[width=85mm]{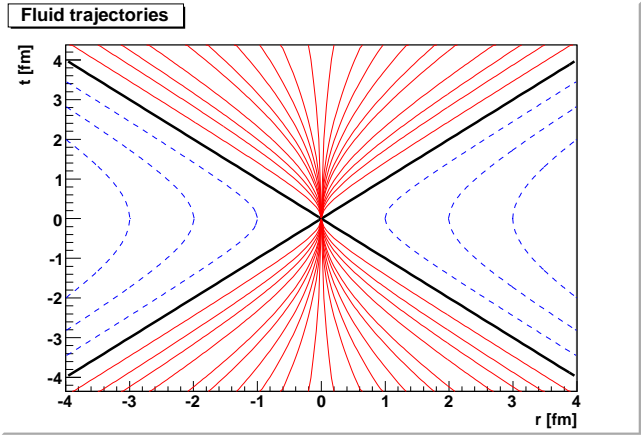}
\caption{Fluid trajectories of the already known~\cite{Nagy:2007xn} exact solution, the $\rho=0$ case of \Eq{e:NCC_mu},
inside (solid) and outside the lightcone (dashed).}
\label{f:trajmu0}
\end{figure}

\begin{figure}
\includegraphics[width=85mm]{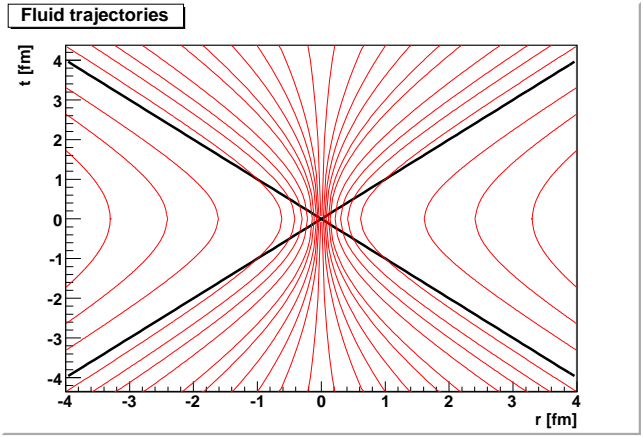}
\caption{Fluid trajectories of the new, accelerating $\eta$-finite exact solution of \Eq{e:NCC_mu}, for the $\rho=1$
case as illustration.}
\label{f:trajmu1}
\end{figure}

The fluid trajectories of the (already known) $\rho=0$ solution are uniformly accelerating in their local rest
frame, Fig. \ref{f:trajmu0} shows these trajectories. Inside the lightcone they describe an explosion from an
initially infinitely dense point at rest. The trajectories of the $\rho>0$ solutions are qualitatively
similar, (shown on Fig.~\ref{f:trajmu1} for $\rho=1$), but they describe an explosion where the initial point
is ,,smeared'' on a distance scale of $\sim\sqrt{\rho}$. The $\v{a}\neq 0$ case describes an explosion where
this initial sphere has nonzero velocity and pressure gradient.

We can have solutions where the initial fireball has nonzero rotation, so has the whole velocity field. Again with
arbitrary constant $\rho$ and constant three-vector $\v{a}$, and a new constant three-vector $\v{B}$, the solution is
\bl{e:FNCC}
\v{v}=\frac{2t\v{r}+\v{a}+\v{B}\times\v{r}}{t^2+r^2+\rho},\quad p=p_0\z{\frac{T}{T_0}}^{\kappa+1} ,
\ee
\bl{e:FNCC2}
T=\frac{T_0\tau_0^2}{\sqrt{\z{\tau^2+\rho}^2+4\rho r^2-\z{\v{B}\times\v{r}+\v{a}}^2-4t\z{\v{a}\v{r}}}} ,
\ee

There exists a rotating generalization of the Hubble-like flow of \Eq{e:BL} as well (for the derivation, see
Appendix \ref{s:app_hydro}):
\bl{e:FHB}
\v{v}=\frac{\v{r}+\v{E}t+\v{B}\times\v{r}}{t+\z{\v{E}\v{r}}},\quad p=p_0\z{\frac{T}{T_0}}^{\kappa+1} ,
\ee
\bl{e:FHB2}
T=\frac{T_0\tau_0}{\sqrt{\tau^2+\z{\v{B}\v{r}}^2-B^2r^2-E^2t^2+\z{\v{E}\v{r}}^2}} ,
\ee
with constant $\v{E}$ and $\v{B}$ three-vectors. (In a special case, if $\v{E}\v{B}=0$, one of the
$\v{E}$, $\v{B}$ vectors can be set to zero.) This flow also describes expansion, where the perturbation with
respect to the Hubble-like flow do not necessarily tend to zero asymptotically. (However, it is \emph{not} only
the Hubble flow seen from a rotating reference frame.)

So far all the solutions presented are Knudsen flows, in the sense used in the introduction. However, if one
does not consider the conserved particle density $n$, only $T$, $p$ and $u^\mu$, one has valid new solutions
(and the already known Hubble-like solution) to the relativistic perfect fluid hydrodynamical equations (not
restricted to collisionless evolution of conserved massless particles). We see that one can have solutions
describing an expansion of an initial fireball that can have finite size, initial velocity and directional
pressure gradient as well as initial rotation. (To my best knowledge, rotating relativistic hydrodynamical
solutions were never found before.) These properties make these solutions themselves an interesting tool in
the description of high energy heavy ion reactions, where a rotating fireball of finite size is initially present. 

If we abandon the requirement of collisionlessness, we can generalize the solutions presented above. For any
velocity field there exists a scalar function $S$ which satisfies $\pd{S}{t}+\v{v}\nabla S=0$. For such $S$,
one can take any $\c{V}\z{S}$ function, and replace $T\to T/\c{V}\z{S}$, $n\to n\cdot\c{V}\z{S}$,
$p\to p$. The obtained formulas solve the relativistic hydrodynamical equations (the only difference would be
in the particle density continuity equation \r{e:n_rel}; it can be checked that the solution remains valid),
but the collisionless Boltzmann equation is solved only if $\c{V}\z{S}\equiv 1$. For example, the $S$ function
for the solution \r{e:NCC_mu} is easily found, and thus the following solution is a finite hydrodynamical
solution with an arbitrary scaling function:
\[
\v{v}=\frac{2t\v{r}}{t^2+r^2+\rho},\quad S=\frac{r}{\tau^2+\rho},\quad
\]
\[
T=T_0\z{\frac{p}{p_0}}^{\rec{\kappa+1}}\rec{\c{V}\z{S}},\quad n=n_0\z{\frac{p}{p_0}}^{\frac{\kappa}{\kappa+1}}\c{V}\z{S} ,
\]
\bl{e:NCC_mu2}
p=\frac{p_0\tau_0^{2\z{\kappa+1}}}{\z{\z{\tau^2+\rho}^2+4\rho\tau^2\sinh^2\eta}^{\frac{\kappa+1}{2}}} ,
\ee
and since $p$ is finite in $\eta$ (as explained after \Eq{e:NCC_mu}), $n$ and $T$ can be finite simultaneously.

\section{Discussion and summary}

In this paper, the connection between hydrodynamical evolution and collisionless kinetic phase-space evolution was
explored. It turns out that in some cases the hydrodynamical flow solves the collisionless Boltzmann
equation as well: thermalization can be maintained by free streaming. This means that in these cases, one can
prepare a special initial condition for the phase-space distribution where the collisionless motion of the particles
maintains local thermalization.

The very early phase of a heavy-ion or hadron collision process, with quarks as degrees of freedom, is a domain
where this ,,Knudsen flow'' may have some implications. With increasing energy the lack of collisions (the
decrease of the cross section due to asymptotic freedom) as well as the ultrarelativistic limit (which, from a
statistical physical point of view, is equivalent to the massless limit) may become realistic, it is interesting to
see that a collective, hydrodynamical motion (even with acceleration) can result from only the initial conditions
for the phase-space distibution, without the need for collisions to achieve the hydrodynamical character. Thus it
seems that in some cases the thermalization process is less connected with the actual collisions than one would think.

The new exact solutions of relativistic perfect fluid hydrodynamics, which emerged during the simultaneous
investigation of the collisionless Boltzmann equation and the hydrodynamical equations, are important on their own right,
and are not necessarily restricted to collisionless evolution of conserved massless particles (only the
$\varepsilon=\kappa p$ equation of state is necessary). Among these solutions, one has accelerating solutions
beyond spherical symmetry, and realistic, accelerating solutions describing expansions with finite space-time rapidity
profile. For the first time, rotating relativistic hydrodynamical solutions were also found. All these new solutions
definitely deserve future work and investigations to establish their connection to the experimentally observable
quantities\footnote{During the finalization of this manuscript I became aware of a different class of new exact
solution of the relativistic hydrodynamical equations that utilizes conformal symmetry in the transverse plane
to generate radial flow~\cite{Gubser:2010ze}. This solution deserves future investigation as well.}.

\section{Acknowledgements}

I would like to thank Tam\'as Cs\"org\H{o} for his help and motivation. This work was supported by the Hungarian
OTKA grants T038406 and T049466.

\appendix

\section{Nonrelativistic collisionlessness}\label{s:app_nonrel}

In this Appendix it is shown that the only reasonable three-dimesional non-relativistic collisionless hydrodynamical
solution is the one discovered in Ref.~\cite{Csizmadia:1998ef}, and recited in Section~\ref{s:notation}. The derivation
goes in the same vein as in the relativistic case in Appendix \ref{s:app_colless}: we expand the collisionless
Boltzmann-equation \r{e:MB_nonrel}, and require that the terms of different order in $\v{p}$ vanish
separately, for all $\v{p}$. From \Eqs{e:f_nonrel}{e:MB_nonrel}, keeping in mind the normalization of \Eq{e:f_nonrel},
we have:
\[
\rec{n}\z{\pd{n}{t}+\frac{\v{p}}{m_0}\nabla n}+\frac{\z{\v{p}-m_0\v{v}}^2}{2m_0T^2}\z{\pd{T}{t}+\frac{\v{p}}{m_0}\nabla T}+
\]
\[
-\frac{3}{2T}\z{\pd{T}{t}+\frac{\v{p}\nabla T}{m_0}}+\frac{\z{\v{p}-m_0\v{v}}}{T}\z{\pd{\v{v}}{t}+\frac{\v{p}\nabla}{m_0}\v{v}}=0 .
\]
We have from the terms of order $p^3$ and $p^2$
\bl{e:homT}
\nabla T = 0 ,\quad \rec{T}\pd{T}{t}\delta_{ik}+\partial_iv_k+\partial_kv_i=0 ,
\ee
which has the unique solution
\bl{e:T_nonrel}
T=T_0\frac{a^2\z{0}}{a^2\z{t}},\quad \v{v}=\frac{\dot{a}\z{t}}{a\z{t}}\v{r}+\v{C}+\v{D}\times\v{r} ,
\ee
with arbitrary $a\z{t}$ function and $\v{C}$, $\v{D}$ constant vectors. Taking the curl of the Euler equation,
\Eq{e:eul_nonrel}, and taking \Eq{e:homT} into account, one gets that if $\dot{a}\z{t}\neq 0$, then $\v{D}=0$.
(The other case is not really interesting.) \Eq{e:T_nonrel} means a solution of the energy continuity equation
\r{e:energy_nonrel} if and only if $\kappa=D/2$. For the moment we assume $\v{C}=0$, in that case the
velocity field is spherically symmetric, and this implies that $p$, and hence $n$ must be spherically
symmetric. With some calculation one can now write down the general solution of the continuity equation
for $n$, \Eq{e:cont_nonrel}. What we have now is
\bl{e:v_nonrel}
T=T_0\frac{a^2\z{0}}{a^2\z{t}},\quad \v{v}=\frac{\dot{a}\z{t}}{a\z{t}}\v{r} ,
\ee
\bl{e:n_nonrel}
n=n_0\frac{a^3\z{0}}{a^3\z{t}}\nu\z{S} ,\quad S\z{t,\v{r}}=\frac{r^2}{a^2\z{t}} .
\ee
The $\nu\z{S}$ function can be arbitrary at this stage. The variable $S$ has the property that
$\pd{S}{t}+\v{v}\nabla S=0$. Now we only have to put these in the Euler equation to obtain the
unique solution as in \Eq{e:sol_nonrel}. It indeed solves the Boltzmann equation, too.
With some calculation one can be convinced that the free $\v{C}$ vector is nothing but the freedom
of a Galilei-transformation: we can observe this flow from a moving reference frame.

\section{Relativistic collisionless evolution}\label{s:app_colless}

The relativistic phase-space distribution $f\z{x,p}$ is given by \Eq{e:f_rel}, for a hydrodynamical flow. To
have a collisionless hydrodynamical flow, besides the equations of hydrodynamics, \Eq{e:MB_rel} also must hold,
for all allowed $p^\mu$ values. So, from \Eqs{e:MB_rel}{e:f_rel}, that the first and the second order terms in
$p^\mu$ vanish separately, we get two equations. The first one is
\bl{e:muT}
\partial_\mu\frac{\mu}{T}=0 , 
\ee
which means that $\mu=const.\cdot T$. This, together with \Eq{e:thermo}, yields the Euler equation in a simpler
form: 
\bl{e:Teul_rel}
Tu^\nu\partial_\nu u^\mu=\z{g^{\mu\rho}-u^\mu u^\rho}\partial_\rho T , 
\ee
containing only the temperature and the four-velocity. Also for the \r{e:energy_rel} energy conservation equation,
with \Eq{e:thermo} and the EoS from \Eq{e:EoS}, we get the alternative form (which is essentially the expression
of entropy conservation for the case of the investigated EoS):
\bl{e:Tenergy_rel}
\partial_\mu\z{T^\kappa u^\mu} = 0 . 
\ee
The other constraint from the Boltzmann equation depends on whether we consider a fluid of massive or
massless particles. For the massive case we have 
\bl{e:tensor0}
\partial_\mu\z{\frac{u_\nu}{T}}+\partial_\nu\z{\frac{u_\mu}{T}}=0 ,
\ee
while, if they are massless (or, in other words, we consider ultrarelativistic particles), the constraint is weaker:
\bl{e:tensor1}
\partial_\mu\z{\frac{u_\nu}{T}}+\partial_\nu\z{\frac{u_\mu}{T}}=\frac{\alpha}{T}g_{\mu\nu} ,
\ee
with a free $\alpha\z{x}$ scalar function. (This weakening is due to the fact that for massless particles
$p^\mu p_\mu=0$.) From now on, we investigate the massless case only. Contracting \Eq{e:tensor1} with $u^\nu$,
and using \Eq{e:Teul_rel}, we find $T\alpha=-2u^\rho\partial_\rho T$, so we have
\bl{e:tensor2} 
\partial_\mu\z{\frac{u_\nu}{T}}+\partial_\nu\z{\frac{u_\mu}{T}}+\frac{2}{T^2}u^\rho\partial_\rho T\cdot g_{\mu\nu} = 0 .
\ee
This equation contains the Euler-equation (we get it by contracting with $u^\nu$), and contracting \Eq{e:tensor2}
with $g^{\mu\nu}$ yields $\partial_\mu \z{T^Du^\mu}=0$, which is essentially the entropy conservation equation,
since we have $\kappa=D$. (This is a reason for using $\kappa=D$ in the relativistic case.) It is clear from the
text before \Eq{e:Tenergy_rel} that this form of the entropy conservation equation is equivalent to the energy
conservation law, \Eq{e:energy_rel}, so we see that \Eq{e:tensor2} is the \emph{only} equation we have to solve.
For this purpose we introduce the four-vector $A^\mu\equiv\z{\varphi,\v{A}}\equiv u^\mu/T$. The three-velocity
is $\v{v}=\v{A}/\varphi$. With this, \Eq{e:tensor2} becomes 
\bl{e:tensor4}
A^\rho A^\sigma\kz{2g_{\mu\nu}\partial_\rho A_\sigma-g_{\rho\sigma}\z{\partial_\mu A_\nu+\partial_\nu A_\mu}}=0 . 
\ee
With some investigation (e.g. writing this equation in non-relativistic notation) we find that \Eq{e:tensor4} is
\emph{equivalent} to the one we get by dropping $A^\rho A^\sigma$:
\bl{e:tensor5}
g_{\mu\nu}\z{\partial_\rho A_\sigma+\partial_\sigma A_\rho}-g_{\rho\sigma}\z{\partial_\mu A_\nu+\partial_\nu A_\mu}=0 . 
\ee
Remarkably, we have obtained a set of linear differential equations. For solving this, let's expand $A_\mu$ in
a Taylor-like series:
\be
A_\mu=a^{(1)}_\mu+a^{(2)}_{\mu\nu}x^\nu+a^{(3)}_{\mu\nu\rho}x^\nu x^\rho+a^{(4)}_{\mu\nu\rho\sigma}x^\nu x^\rho x^\sigma+\dots .
\ee
The $a^{(i)}$ coefficient tensors are by definition totally symmetric in their last $i-1$ indices. This expansion
works if both $T$ and $u^\mu$ are smooth, and $T\neq 0$. (We should have such domain in space-time if we are after
hydrodynamical solutions.) From the requirement that \Eq{e:tensor5} holds order by order in $x^\mu$, we have
\[
g_{\rho\sigma}a^{(2)}_{\sz{\mu\nu}}       =g_{\mu\nu}a^{(2)}_{\sz{\rho\sigma}}        ,\quad
g_{\rho\sigma}a^{(3)}_{\sz{\mu\nu}\lambda}=g_{\mu\nu}a^{(3)}_{\sz{\rho\sigma}\lambda} ,\quad
\]
\bl{e:tensor6}
g_{\rho\sigma}a^{(4)}_{\sz{\mu\nu}\lambda\eta}=g_{\mu\nu}a^{(4)}_{\sz{\rho\sigma}\lambda\eta} ,\quad\dots ,
\ee
where square brackets mean symmetrization. To proceed, we must now distinguish between the $D=1$ and the $D\neq 1$ case.

In the $D\neq 1$ case, \Eq{e:tensor6} is satisfied for $a^{(2)}_{\mu\nu}$ if
\[
a^{(2)}_{\mu\nu}=\gamma g_{\mu\nu}+F_{\mu\nu} ,
\]
with arbitrary given $\gamma$ constant, $b^\mu$ vector and $F^{\mu\nu}$ antisymmetric tensor. The equations for
the $a^{(i)}$s for $i>2$ imply
\bl{e:aT}
a^{(i)}_{\mu\nu\lambda\dots}+a^{(i)}_{\nu\mu\lambda\dots}=g_{\mu\nu}T^{(i)}_{\lambda\dots} ,
\ee
with totally symmetric $T_{\lambda\dots}$ tensors. Taking the antisymmetric part of this equation in $\mu$ and
$\lambda$, and then in $\mu$ and $\nu$, using the symmetry properties of the $a^{(i)}$s, and again \Eq{e:aT}, we obtain 
\be
a^{(i)}_{\mu\nu\lambda\dots}=\rec{2}\z{g_{\mu\nu}T^{(i)}_{\lambda\dots}+g_{\mu\lambda}T^{(i)}_{\nu\dots}-g_{\nu\lambda}T^{(i)}_{\mu\dots}} .
\ee
For $i=3$, taking $T^{(3)}_\lambda=-b_\lambda$, we immediately have the result as
\bl{e:a2a3}
a^{(3)}_{\mu\nu\lambda}=\rec{2}\z{g_{\mu\lambda}b_{\nu}+g_{\mu\nu}b_{\lambda}-g_{\nu\lambda}b_\mu} 
\ee
For $i>3$, that $a^{(i)}_{\mu\nu\lambda\eta\dots}$
be symmetric in $\lambda\leftrightarrow\eta$, we can express these constraints as
\[
\z{g_{\mu\nu}T^{(i)}_{\lambda\eta\dots}-g_{\lambda\nu}T^{(i)}_{\mu\eta\dots}}-
\z{g_{\mu\eta}T^{(i)}_{\lambda\nu\dots}-g_{\lambda\eta}T^{(i)}_{\mu\nu\dots}}=0 .
\]
For every set of the other remaining indices (denoted collectively by dots) this homogeneous linear system of
equations has $D\z{D+1}\z{D\z{D+1}+2}/8$ independent components (because the antisymmetry in
$\mu\leftrightarrow\lambda$ and $\nu\leftrightarrow\eta$, and the symmetry in $\mu\lambda\leftrightarrow\nu\eta$),
while a symmetric tensor $T_{\mu\nu}$ has $\z{D+1}\z{D+2}/2$ independent components. So for $D=2$ there are $6$
equations for $6$ unknowns, for $D\ge 3$ there are more equations than unknowns. Furthermore, this set of linear
equations is nonsingular, so there can be no nontrivial solutions for $i\ge 4$ and $D\ge 2$. Combining these
together, we arrive at the general solution for \Eq{e:tensor5} as
\bl{e:Amu}
A^\mu=a^\mu+\gamma x^\mu+F^{\mu\nu}x_\nu+\z{b^\nu x_\nu}x^\mu-\frac{\z{x^\nu x_\nu}b^\mu}{2} ,
\ee
where $a_\mu\equiv a^{(1)}_\mu$ is an arbitary four-vector. From this, we get the solutions derived in
Appendix \ref{s:app_hydro} and described in Section \ref{s:hydro}.

The $D=1$ case case is simpler: \Eq{e:tensor5} yields only the following conditions for $\varphi\,\z{\equiv A^0}$
and $A\equiv A^1$:
\be
\pd{A}{t}=\pd{\varphi}{x} ,\quad \pd{A}{x}=\pd{\varphi}{t} .
\ee
(Here $x\equiv x^1$ is the only spatial coordinate.) Again with the notation $x^\pm\equiv t\pm x$, this has the
following general solution, with arbitrary $\chi\z{x^+}$ and $\xi\z{x^-}$ functions:
\bl{e:1Dsol}
A=\frac{\chi\z{x^+}-\xi\z{x^-}}{T_0} ,\quad \varphi =\frac{\chi\z{x^+}+\xi\z{x^-}}{T_0} ,
\ee
where an arbitrary temperature scale $T_0$ is introduced. Having the definition of $A^\mu$ in mind, we arrive at
the solution as in \Eq{e:1DvT}.

We thus found \emph{all} those relativistic hydrodynamical solutions, where thermalization can be maintained without
the collision of particles (in the case of massless particles).

\section{Three-dimensional collisionless flows}\label{s:app_hydro}

This Appendix investigates the solutions stemming from \Eq{e:Amu}. We use the Rindler-like coordinates
introduced in \Eq{e:Rindler}.
\begin{itemize}
\item
If the $b^\mu$ vector and the $F^{\mu\nu}$ tensor in \Eq{e:Amu} is zero, then, if $\gamma=0$, the solution is
a medium at rest, but if $\gamma\neq 0$, then $a^\mu$ can be eliminated with a translation
$x^\mu\to x^\mu-a^\mu/\gamma$, and introducing the notation $T_0\tau_0\equiv1/\gamma$ (to set the scale of
the initial temperature, we obtain the solution cited in \Eq{e:BL}, the Hubble-like expanding solution. 
\item
We can have $b^\mu\neq 0$ in \Eq{e:Amu}, and we can redefine $F^{\mu\nu}$ and $a^\mu$ in \Eq{e:Amu} and
make a translation in $x^\mu$ to make $\gamma$ vanish. The simplest case then is when $b^\mu$ is time-like and
$F^{\mu\nu}=0$: we can adopt a frame where $b^\mu=\z{\zeta,\v{0}}$, and we obtain a new accelerating solution
(with $a^\mu\equiv\z{\zeta\rho/2,\zeta\v{a}/2}$ and $2/\zeta\equiv T_0\tau_0^2$) as in \Eq{e:NCC_mu}.
\item
For time-like $b^\mu$ and nonvanishing $F_{\mu\nu}$ we obtain accelerating flows with curl. We can choose a
frame where $b^\mu=\z{\zeta,\v{0}}$, and adopt the three-dimensional notation for the $F_{\mu\nu}$ tensor
as $F_{0i}=\rec{2}\zeta E_i$, $F_{ik}=\rec{2}\zeta\epsilon_{ikl}B_l$, take $a^\mu\equiv\z{\zeta\rho/2,\zeta\v{a}/2}$.
It turns out that with an appropriate translation and redefinition of the other constants, we can eliminate
$\v{E}$. We thus have the solution as in \Eqs{e:FNCC}{e:FNCC2}, with $T_0\tau_0^2=2/\zeta$.
\item
The rotating generalizations of the Hubble-flow, given in \Eqs{e:FHB}{e:FHB2} can be derived from \Eq{e:Amu}
if we take $b^\mu=0$, $\gamma\neq 0$, so we have $A^\mu=\gamma x^\mu+F^{\mu\nu}x_\nu$. With the notation
$F_{0i}=\gamma E_i$, $F_{ik}=\gamma\epsilon_{ikl}B_l$, the solution is given by \Eqs{e:FHB}{e:FHB2}
(with $T_0\tau_0\gamma=1$).
\item
For space-like $b^\mu$, we can choose a reference frame where $b^\mu=\z{0,\v{b}}$, and obtain collisionless ,,solutions''.
However, they are not physical, since the velocity of some initially slow fluid trajectories will approach and surpass the speed
of light.
\end{itemize}

\end{document}